\documentclass[aps,prl,floatfix,superscriptaddress,twocolumn,footinbib]{revtex4-2}

\usepackage{enumitem}
\usepackage{amssymb}
\usepackage{amsmath}
\usepackage{amsfonts}
\usepackage{appendix}
\usepackage{bm}
\usepackage{balance}
\usepackage{booktabs}
\usepackage{braket}
\usepackage{calc}
\usepackage{epsfig}
\usepackage{epstopdf}
\usepackage{graphicx}
\usepackage{lipsum}
\usepackage{natbib}
\usepackage[dvipsnames]{xcolor}

\usepackage[colorlinks,linkcolor=blue,anchorcolor=blue,citecolor=blue,urlcolor=blue]{hyperref}

\begin{document}

\title
{
Theory of Correlated Hofstadter Spectrum in Magic-Angle Graphene 
}

\author{Chen Zhao}
\affiliation{School of Physics and Wuhan National High Magnetic Field Center, Huazhong University of Science and Technology, Wuhan, Hubei 430074, China}

\author{Zhaowen Miao}
\affiliation{School of Physics and Wuhan National High Magnetic Field Center, Huazhong University of Science and Technology, Wuhan, Hubei 430074, China}
\author{Zhen Ma}
\affiliation{School of Electronics and Information, Zhengzhou University of Light Industry, Zhengzhou, 450002 China}

\author{Ying-Hai Wu}
\email{yinghaiwu88@hust.edu.cn}
\affiliation{School of Physics and Wuhan National High Magnetic Field Center, Huazhong University of Science and Technology, Wuhan, Hubei 430074, China}    

\author{Ming Lu}
\email{luming@baqis.ac.cn}
\affiliation{Beijing Academy of Quantum Information Sciences, Beijing 100193, China}

\author{Jin-Hua Gao}
\email{jinhua@hust.edu.cn}
\affiliation{School of Physics and Wuhan National High Magnetic Field Center, Huazhong University of Science and Technology, Wuhan, Hubei 430074, China}

\author{X. C. Xie}
  \affiliation{International Center for Quantum Materials, School of Physics, Peking University, Beijing 100871, China}
  \affiliation{Institute for Nanoelectronic Devices and Quantum Computing, Fudan University, Shanghai 200433, China}
  \affiliation{Hefei National Laboratory, Hefei 230088, China}

\begin{abstract}
The magnetic-field-induced correlated Chern insulator (CCI) states in magic-angle twisted bilayer graphene (MATBG) have been intensively studied in experiments, but a simple and clear understanding of their origin is still lacking. Here, we propose a unified theoretical framework for the CCI states in MATBG that successfully explains most experimental observations. The key insight of our theory is that, due to the very narrow bandwidth of MATBG, correlation-enhanced valley and spin Zeeman terms are critical for shaping the intricate Hofstadter spectrum, resulting in an interwoven, flavor-resolved (spin and valley) Hofstadter spectrum that can well describe the observed CCI states. Crucially, due to the Zeeman effect, the crossings between these flavor-polarized Hofstadter spectra are magnetic-field-dependent, causing certain CCI states to emerge only above a critical field. This is the main mechanism underlying the critical field phenomenon of the CCI states observed in experiments. Our theory provides a clear and unified physical picture for the correlated Hofstadter spectrum in MATBG.
\end{abstract}

\maketitle

\emph{Introduction.} ---The Hofstadter butterfly represents an intriguing fractal energy spectrum that arises in a two-dimensional electron gas subjected simultaneously to a perpendicular magnetic field and a periodic lattice potential~\cite{hofstadter1976energy}. A key characteristic of this spectrum is the emergence of Chern insulators when the Fermi energy lies within a gap, which are indexed by a pair of integers ($t$, $s$), with $t$ representing the Chern number and $s$ the band filling factor~\cite{wannier1978result,thouless1982quantized,KOHMOTO1985343,IDana_1985,kohmoto1989zero}.

The Hofstadter spectrum in magic-angle twisted bilayer graphene (MATBG) is of particular interest because it provides a unique platform for studying strongly correlated Hofstadter physics~\cite{bistritzer2011moire,hejaz2019landau,moon2012energy,zhangyahui2019landau,Moon2013optical,Hasegawa2013periodic,Wang2024theory,Wania2024atomistic,Guan2022reentrant,Chou2020minimally,cao2018correlated,tomarken2019electronic,yankowitz2019tuning,sameer2022chern,uri2020mapping,xie2019spectroscopic,lu2019superconductors,cheng2021emergence,joe2022unusual,xiaobo2021multiple,schmidt2014superlattice,nuckolls2025spectroscopy,park2021flavour,nuckolls2020strongly,choi2021correlation,wu2021chern,hu2025link,yu2022correlated,saito2021hofstadter,das2021symmetry,he2025strongly,PhysRevLett.127.197701,pierce2021unconventional,polski2022hierarchy,xie2021fractional}. The magic-angle condition creates a large moiré supercell, allowing the spectrum to be observed at low magnetic fields.  Meanwhile, the extremely narrow bandwidth enhances the Coulomb interactions. Together with the spin and valley degrees of freedom, these conditions enable MATBG to host exotic topological correlated states, revealing a subtle interplay between topology and correlation in these Hofstadter subbands.

In experiments, the strongly correlated Hofstadter spectrum in MATBG has been intensively studied using multiple measurement techniques~\cite{cao2018correlated,tomarken2019electronic,yankowitz2019tuning,sameer2022chern,uri2020mapping,xie2019spectroscopic,lu2019superconductors,cheng2021emergence,joe2022unusual,xiaobo2021multiple,schmidt2014superlattice,nuckolls2025spectroscopy,park2021flavour,nuckolls2020strongly,choi2021correlation,wu2021chern,hu2025link,yu2022correlated,saito2021hofstadter,das2021symmetry,he2025strongly,PhysRevLett.127.197701,pierce2021unconventional,polski2022hierarchy,xie2021fractional}. A hallmark of this spectrum is the emergence of a series of interaction-induced, flavor-polarized (spin and valley) correlated Chern insulator (CCI) states~\cite{park2021flavour,nuckolls2020strongly,choi2021correlation,wu2021chern,hu2025link}. Furthermore, state-of-the-art measurements have revealed even more exotic phenomena, including symmetry-broken Chern insulators~\cite{yu2022correlated,saito2021hofstadter,das2021symmetry,he2025strongly,PhysRevLett.127.197701,pierce2021unconventional,polski2022hierarchy}—which break the discrete translational symmetry of the moiré lattice—as well as fractional Chern insulator states~\cite{xie2021fractional,he2025strongly}. However, a comprehensive theoretical understanding of the correlated Hofstadter spectrum of MATBG is still rare~\cite{Wang2024theory}, even for the basic CCI states. 

In this work, we establish a unified theoretical framework for the correlated Hofstadter spectrum of MATBG,  which provides a comprehensive explanation for nearly all experimentally observed phenomena associated with CCIs.
The key insight of our theory is that, due to the extremely narrow bandwidth of MATBG, the Zeeman splittings of spin and valley---when accounting for the orbital magnetism and the enhancement of the $g$-factor by correlation effects~\cite{lu2019superconductors,jiang2019valley,vallejo2021detection,PhysRevB.102.121406,du2009fractional,song2010high,jeong2024interplay,PhysRevLett.96.136806,PhysRevLett.108.106804,PhysRevLett.124.097601,PhysRevLett.99.106802,PhysRevB.100.085437,ren2020spectroscopic}---play a critical role in determining the fine structure of the Hofstadter spectrum. By incorporating the correlation-enhanced Zeeman splitting, we finally derive a set of interwoven,  \textit{flavor}-polarized Hofstadter spectra for MATBG that can well describe the observed CCI states. Crucially, the Zeeman effect renders the crossings between \textit{flavor}-polarized Hofstadter spectra magnetic-field-dependent, causing certain CCI states to emerge only when the magnetic field exceeds a critical value. This explains the origin of the experimentally observed critical fields.

\begin{figure*}[ht!]
\centering
\includegraphics[width=18cm]{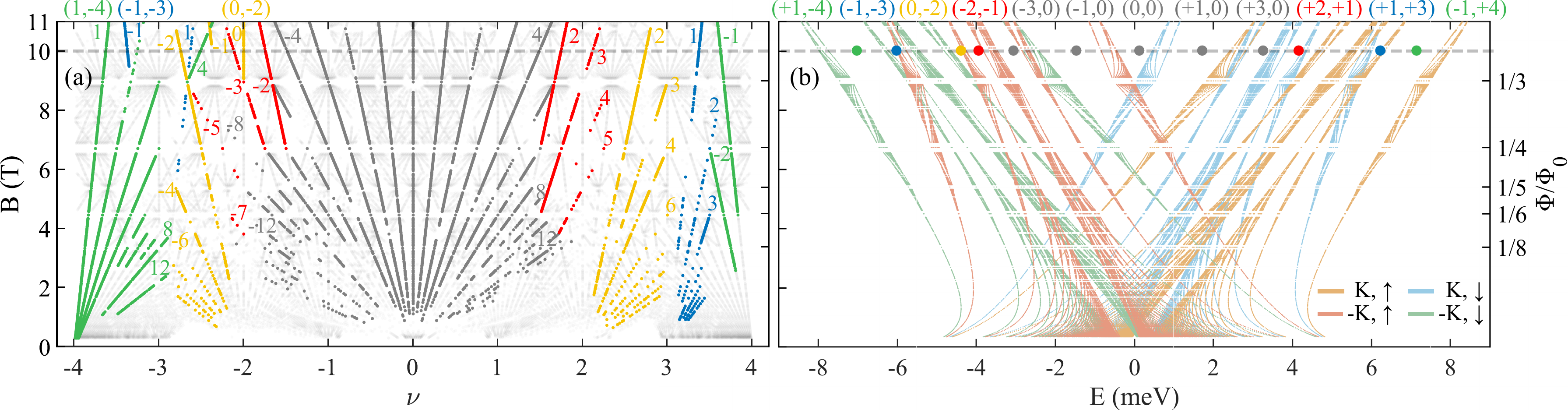}
\caption{Hofstadter spectrum of  MATBG with twist angle $\theta=1.06^\circ$. (a) Wannier diagram. (b) Flavor-polarized Hofstadter spectra. Each gap (CCI state) in Hofstadter spectrum corresponds to a linear trajectory in Wannier diagram denoted by two integers $(t,s)$, where $t$ is the Chern number and $s$ represents the band filling. In Wannier diagram, the observed CCI states in experiment are colored according to their value of $s$, with the same color scheme as that in Ref.~\cite{yu2022correlated}. The integers represent the value of $t$ for these CCI states. The minimum energy resolution for gaps is $0.01$ meV. }
\label{fig1}
\end{figure*}

This physical picture rests on two key assumptions:
\begin{enumerate}[label=(\arabic*), leftmargin=*, nosep, before=\vspace{0.5\baselineskip}, after=\vspace{0.5\baselineskip}]
      \item In MATBG, correlation effects renormalize and enhance the $g$-factors for both spin and valley. Through comparison with experimental data, we determine the renormalized $g$-factors to be $g_s \approx 3.4$ for spin and $g_v \approx 7.7$ for valley.

     \item A non-aligned hBN substrate still induces a small mass term that breaks sublattice symmetry, lifting valley degeneracy and endowing the two valleys with distinct orbital magnetic moments, thereby generating valley Zeeman splitting~\cite{PhysRevB.81.195431,wong2023insulators,hunt2013massive,lu2019superconductors,long2022atomistic,PhysRevB.102.155136,PhysRevB.76.073103,PhysRevB.107.115140,PhysRevB.96.085442,jung2015origin}. Such tiny mass term does not induce a significant transport gap.
\end{enumerate}
Based on these assumptions, we compute the correlated Hofstadter spectrum for MATBG and perform detailed comparisons with multiple experiments. Finally, we predict that applying an \textit{in-plane} magnetic field can tune the intersection of the Hofstadter spectra via the spin Zeeman effect, thereby controlling the behaviors of CCI states. This prediction serves as a critical test to further validate our theory.

\emph{Model and Methods.}---We consider TBG at a twist angle  $\theta$ in a uniform magnetic field
 $\mathbf{B}=(0, 0, B_z)$ perpendicular to the graphene layers.  The Hamiltonian is
\begin{equation}\label{eq1}
    H = H_0+H_\mathrm{V}+H_\mathrm{S}
\end{equation}
where $H_0= H_{\mathrm{intra}} + H_{\mathrm{inter}}$ is the BM model of MATBG in a magnetic field~\cite{BMmodel,bistritzer2011moire,hejaz2019landau}. The intralayer part of $H_0$ is
\begin{equation}\label{CM}
\begin{aligned}
    H_{\mathrm{intra}} = &v_F [(p_x+A_x-\hbar K_x)\tau_z\sigma_x \\
    &+(p_y+A_y-\tau_z\eta_z\hbar K_y)\sigma_y]+\Delta\sigma_z
\end{aligned},
\end{equation}
where $\tau_i,\eta_i$, and $\sigma_i$ are Pauli matrices acting on the valley, layer, and sublattice subspaces, respectively. $\mathbf{K}=k_\theta(\frac{\sqrt{3}}{2},-\frac{1}{2})$ represents the shift of one Dirac point relative to another by the rotation. $k_\theta=2k_\text{D}\sin{(\theta/2)}\approx k_\text{D}\theta$ with $k_\text{D}$ being the Dirac momentum. $\Delta\sigma_z$ is a gap-opening term induced by substrates.  
The interlayer part is 
\begin{equation}
    H_{\mathrm{inter}}=\eta^+T(\mathbf{r})+\mathrm{H.c.}
\end{equation}
with $\eta^+=\frac{1}{2}(\eta_x+i\eta_y)$. $T(\mathbf{r})=\sum_{n=0}^2T_ne^{-i\tau_z\mathbf{q}_n\cdot\mathbf{r}}$, where $T_n=t_{\textrm{AA}}+t_{\textrm{AB}}[\cos(n\phi)\sigma_x+\sin(n\phi)\tau_z\sigma_y]$. Here $\mathbf{q}_0=(0, 0), \mathbf{q}_1=\frac{\sqrt{3}}{2}k_\theta(-1, \sqrt{3}), \mathbf{q}_2=\frac{\sqrt{3}}{2}k_\theta(1, \sqrt{3})$, and $\phi=2\pi/3$.  We set $t_{\textrm{AA}}=0.08$ eV and $t_{\textrm{AB}}=0.1$ eV, for which the first magic angle occurs at $\theta \approx 1.12^\circ$~\cite{MA202118,PhysRevX.8.031087}.

$H_\mathrm{V}$ and  $H_\mathrm{S}$ are the valley and spin Zeeman terms 
\begin{equation}
    H_\mathrm{V}+H_\mathrm{S} = g_\mathrm{v}\mu_BB_z\tau_z+\frac{1}{2}g_\mathrm{s}\mu_BB_zs_z,
\end{equation}
where $\mu_B$ is the Bohr magneton, and  $s_i$ $(i=x,y,z)$ is the Pauli matrices acting on the spin subspaces. $g_\mathrm{v}$ and $g_\mathrm{s}$ are the effective g-factors of valley and spin splitting, respectively. 
The Zeeman terms play a key role in shaping the Hofstadter spectrum in MATBG. 
In monolayer graphene and TBG, magnetic-field-induced valley and spin splittings have been extensively studied in experiments, which can be approximately described by valley and spin Zeeman terms with correlation-enhanced  $g$-factors~\cite{lu2019superconductors,jiang2019valley,vallejo2021detection,PhysRevB.102.121406,du2009fractional,song2010high,jeong2024interplay,PhysRevLett.96.136806,PhysRevLett.108.106804,PhysRevLett.124.097601,PhysRevLett.99.106802,PhysRevB.100.085437,ren2020spectroscopic}.   In MATBG, the narrow bandwidth amplifies these effects to dominate the Hofstadter spectrum. Specifically,  the valley and spin Zeeman effects inherently induce a flavor-polarized Hofstadter spectrum. Consequently, at experimentally accessible magnetic fields, the Hofstadter spectrum of MATBG ultimately manifests as the interweaving of the four flavor-polarized butterfly spectra. The numerical calculations reveal that such an intricate crossing pattern is the primary mechanism underlying the CCI states observed in experiments. 

The other key assumption is the substrate-induced mass term. In graphene systems, the valley Zeeman effect is generally understood to originate from the valley-contrasting orbital magnetic moment~\cite{PhysRevB.81.195431}. To account for this, we introduce a small mass term $\Delta$ in $H_0$, induced by the hBN substrate, to break the sublattice symmetry. This term generates the requisite valley-contrasting orbital magnetic moment without introducing a significant transport gap. Our numerical results confirm that this mass term is vital to reproduce the experimental Hofstadter spectrum in MATBG~\footnote{See Supplemental Material at [URL] for the band structure of MATBG with substrate-induced gaps, the calculation details, the influence of the mass term, the additional examples for the two gap-closing mechanisms, and the comparison of the experimental results.}.

\begin{figure*}
\centering
\includegraphics[width=18cm]{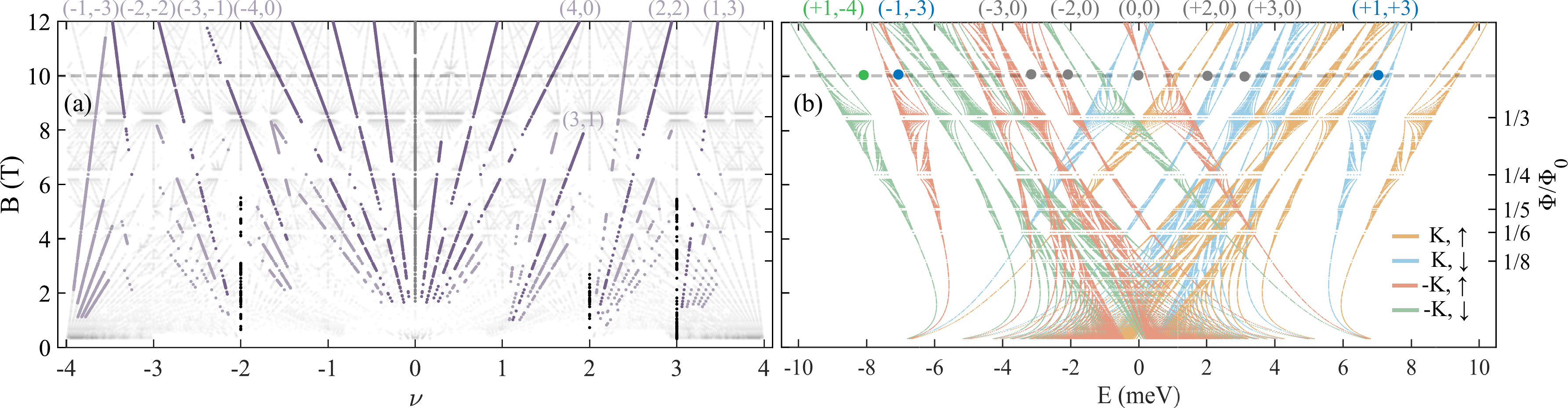}
\caption{Hofstadter spectrum of  MATBG with twist angle $\theta=1.03^\circ$. (a) Wannier diagram. The observed CCI states in experiment are highlighted with the same color scheme as that in Ref.~\cite{he2025strongly}. (b) Flavor-polarized Hofstadter spectra. The minimum energy resolution for gaps is $0.02$ meV. All the other parameters are the same as Fig.~\ref{fig1}.}
\label{fig2}
\end{figure*}

To calculate the Hofstadter spectrum, we use the Landau gauge $\mathbf{A}=B_z(-y,0)$ and express $H_0$ in the Landau level basis $\ket{l n \sigma y_c}$ of graphene monolayer, where $l$ denotes layer, $n$ is the Landau level index and $y_c$ is the guiding center coordinate~\cite{bistritzer2011moire,pfannkuche1992theory,hejaz2019landau}. $ H_\mathrm{intra}$ writes as 
\begin{equation}
    H_\mathrm{intra}=\hbar v_F\tau_z\frac{(\eta_x+i\tau_z\eta_y)}{2}(\frac{\sqrt{2}}{\ell_B}a+i\eta_z\frac{k_\theta}{2})+\mathrm{H.c.}+\Delta\sigma_z
\end{equation}
The matrix element of $H_\mathrm{inter}$ can be calculated with the formula
\begin{equation}\label{eq6}
\begin{aligned}
    \bra{l n \sigma y_c}T_n&e^{-i\mathbf{q}_n\cdot\mathbf{r}}\ket{l^\prime n^\prime \sigma^\prime y_c^\prime}=\delta_{y_c,y_c^\prime}\delta_{k_x+q_{nx},k_x^\prime}F_{n,n^\prime}(z) \\
    &\times\bra{l,\sigma}T_n\ket{l^\prime,\sigma^\prime}e^{-ik_xq_{ny}l^2_B}e^{-iq_{nx}q_{ny}l^2_B/2} 
\end{aligned}
\end{equation}
where $z=(q_{nx}+iq_{ny})\ell_B/\sqrt{2}$, and $F_{n,n^\prime}(z)$ is
\begin{equation}\label{eq7}
    F_{n,n^\prime}(z)=\left\{
\begin{aligned}
    &e^{-zz^\ast/2}z^{n^\prime-n}\sqrt{\frac{n!}{n^\prime!}}L^{n^\prime-n}_n(zz^\ast), &n\le n^\prime \\
    &e^{-zz^\ast/2}(-z^\ast)^{n-n^\prime}\sqrt{\frac{n^\prime!}{n!}}L^{n-n^\prime}_{n^\prime}(zz^\ast), &n> n^\prime
\end{aligned}
\right.
\end{equation}
where $L^a_b(x)$ is the associated Laguerre polynomial. The calculation details are given in the supplementary materials.
We remind the effects of the mass term $\Delta\hat{\sigma}_z$.  Under a magnetic field, this mass term induces opposite shifts of the zeroth Landau level in the two valleys~\cite{PhysRevB.81.195431}, thereby lifting the valley degeneracy and, in turn, producing pronounced reconstructions of the Hofstadter spectra (See Supplemental Material). Correlation effects further enhance the valley and spin splittings, which can be effectively described by renormalized $g$-factors.

\emph{Correlated Hofstadter spectrum.}---The Hofstadter spectrum of MATBG can be viewed as magnetic subbands formed by coupling the Landau levels (LLs) of two graphene monolayers through moir\'e interlayer hopping. Experimentally, the Hofstadter spectrum is determined by measuring the energy gaps between magnetic subbands (or LLs). These gaps manifest as linear trajectories in Wannier diagram described by the Diophantine equation $\nu=t ( \Phi /\Phi_0 )+s $.  $\nu$ and $\Phi$ are the electron filling and magnetic flux per moir\'e unit cell, respectively.  When $E_F$ is in such a gap, a linear trajectory with nonzero $t$  indicates the emergence of a CCI state.

\begin{figure*}
\centering
\includegraphics[width=18cm]{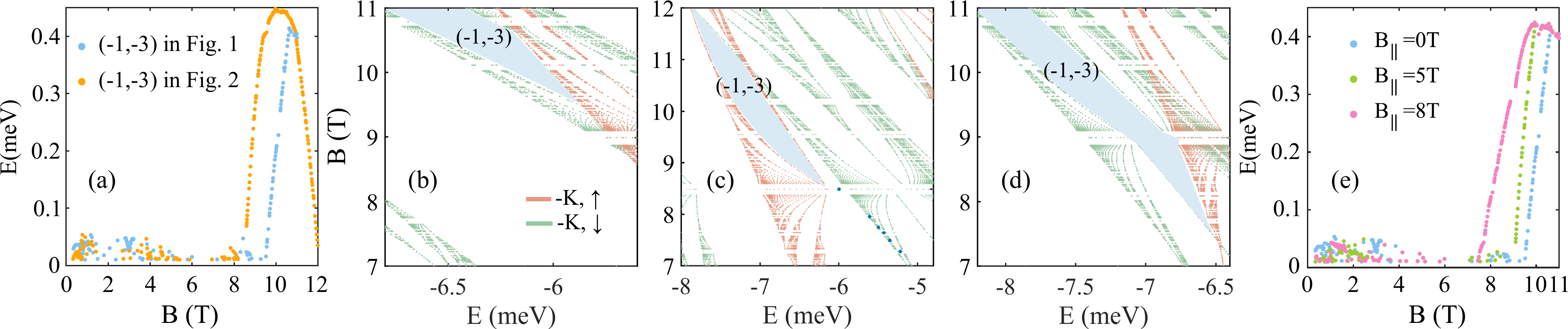}
\caption{(a) is the evolution of the gaps of the $(-1,-3)$ states in both Fig.~\ref{fig1} and Fig.~\ref{fig2} with magnetic field. (b) and (c) are the magnified versions of the Hofstadter spectrum with $\theta=1.06^\circ$ and $\theta=1.03^\circ$, respectively. The blue shaded region marks the gap of the $(-1,-3)$ state. (d) is the magnified version of the Hofstadter spectrum with $\theta=1.06^\circ$ and an in-plane magnetic field $B_\parallel=8$ T. (e) is the evolution of gaps of the $(-1,-3)$ states with $\theta=1.06^\circ$, under different in-plane fields.}
\label{fig3}
\end{figure*}

Although different groups use samples with slightly varying twist angles (all near the magic angle) and employ a variety of techniques---including transport measurements~\cite{yankowitz2019tuning,sameer2022chern,uri2020mapping,lu2019superconductors,cheng2021emergence,joe2022unusual,xiaobo2021multiple,schmidt2014superlattice,park2021flavour,wu2021chern,saito2021hofstadter,das2021symmetry,he2025strongly,PhysRevLett.127.197701,polski2022hierarchy}, scanning tunneling microscopy (STM)~\cite{xie2019spectroscopic,nuckolls2025spectroscopy,nuckolls2020strongly,choi2021correlation}, single-electron tunneling spectroscopy (SET)~\cite{yu2022correlated,pierce2021unconventional,xie2021fractional}, and capacitance measurements~\cite{cao2018correlated,tomarken2019electronic,park2021flavour,hu2025link}---the Hofstadter spectrum consistently exhibits similar features. The most salient one is the emergence of a series of CCI states with integer  $(t, s)$ ($|s| \leq 4$). These CCI states are generally believed to originate from flavor-polarized magnetic subbands, in which interactions lift the spin and valley degeneracy. Notably, some of these CCI states emerge abruptly only when the magnetic field exceeds a critical value $B_{\text{c}}$~\cite{saito2021hofstadter,das2021symmetry,yu2022correlated,xie2021fractional,hu2025link}.

Our Hofstadter spectrum model provides a clear explanation of the experimentally observed CCI states. Here, with a set of effective parameters: $g_s \approx 3.4$, $g_v \approx 7.7$, and $\Delta \approx 1$ meV, we calculate the Hofstadter spectrum, and compare it with a series of experiments~\cite{yu2022correlated,he2025strongly,PhysRevLett.127.197701,saito2021hofstadter,wu2021chern,das2021symmetry}. 

The first experiment is the measurement of the SET at a twist angle  $\theta \approx 1.06^\circ$~\cite{yu2022correlated}. Owing to its nanoscale resolution, which significantly reduces the impact of spatial inhomogeneity, this experiment reveals more details of the interacting Hofstadter spectrum in MATBG. The calculated Wannier diagram and  Hofstadter spectra are presented in Fig.~\ref{fig1}(a) and (b), respectively. First of all, this Wannier diagram accurately reproduces all the observed CCI states in experiment, which are highlighted with different colors according to their $s$ values.  The Wannier diagram can be understood from the  Hofstadter spectrum in Fig.~\ref{fig1} (b). Due to the Zeeman terms, the Hofstadter spectrum here is inherently flavor-polarized. At low magnetic fields, the intersection between flavor-polarized Hofstadter spectra completely destroys the Hofstadter gaps, so that no CCI states with $|s|<4$ are observed in the Wannier diagram. Only in the low-energy region, where spectral crossings are absent, can CCI states be observed under low magnetic fields---as exemplified by the $(t,s)=(1, -4)$ state (see Fig.~\ref{fig1}). As $B$ increases, the intersections between Hofstadter spectra become less severe, allowing the gaps to be resolved experimentally. It is the main reason why some CCI states can only be observed above a critical field. 
The $(-1,-3)$ state is a representative example, corresponding to an intersection point of the Hofstadter spectra near $B\approx 9.5$ T. In short, the spectral crossings induced by Zeeman splitting lead to gap closure, disrupting the linear trajectories---a phenomenon we term the Zeeman-splitting-induced gap-closing mechanism.

There is another general gap-closing mechanism. 
When the magnetic flux takes a rational value $\Phi / \Phi_0 = 1/q$ ($q$ is an integer), the system recovers magnetic translational symmetry, and the fractal energy spectrum and related gaps disappear. The corresponding magnetic field is $B_{1/q}\approx12q(\theta^\circ)^2$~\cite{hejaz2019landau}. As a result, the linear trajectories from the Hofstadter subbands are highly susceptible to interruption at such magnetic fields, while other gaps from Landau levels, e.g., the $(+1,-4)$ state, remain unaffected.  Note that this gap-closing mechanism is distinct from the one stemming from the Zeeman splitting.  As $B$ continues to change, some self-similar linear trajectories may reappear under the effective magnetic field $B_{\textrm{eff}}=B-B_{1/q}$, indicating that the gaps (linear trajectories) intersect at $\Phi/\Phi_0=1/q$~\cite{hofstadter1976energy,wannier1978result,bistritzer2011moire,nuckolls2025spectroscopy}. Compared with the experiment, the crossing of CCI states at $\Phi/\Phi_0=1/3$ ($9$~T), and $\nu=-\frac{8}{3}$ should be a typical example. 

Most of the linear trajectory evolution in Wannier diagram can be explained by the two mechanisms. For example, the $(3, -4)$ state terminates near $\Phi/\Phi_0 = 1/3$ ($9$~T), and the $(-2, -1)$ and $(2, 1)$ states are interrupted near $\Phi/\Phi_0 = 1/4$ ($6.7$~T). They are the gap-closing near rational flux. In contrast, if we carefully examine the details of the Hofstadter spectrum, the termination of the $(0,-2)$ state near $\Phi/\Phi_0 = 1/3$ is actually determined by Zeeman splitting (See Supplemental Material).

The Hofstadter spectrum is highly sensitive to the twist angle of the sample. This can explain why different experiments on similar samples can report markedly different features. Here, we compare with another transport experiment with a sample with $\theta \approx 1.03^\circ$~\cite{he2025strongly}. 
It can be observed that even a slight variation in the twist angle leads to noticeable experimental differences. In Fig.~\ref{fig2} (a) and (b),  we present the corresponding Wannier diagram and Hofstadter spectrum, respectively. Apart from the different $\theta$, all the other parameters remain the same. To facilitate comparison with experimental results, in the calculated Wannier diagram, we highlight the experimentally measured gaps following the color scheme used in the experimental paper. 
First, under our parameters, the $1.03^\circ$ sample exhibits a larger bandwidth of moiré flat bands compared to the $1.06^\circ$ sample (See Supplemental Material). Consequently, in Fig.~\ref{fig2} (b), the Hofstadter bands display significantly broader bandwidths, which leads to a completely different crossing pattern for the Hofstadter spectra. It will substantially affect the experimental results.

We again take the $(-1, -3)$ state as an example, since its Hofstadter spectrum intersection is simple and clear.  
 In Fig.~\ref{fig3} (b) and (c), we present the magnified Hofstadter spectra for these two samples, respectively. The gaps corresponding to the $(-1, -3)$ state are highlighted (blue shaded regions).  In Fig.~\ref{fig3} (a), we further plot the $(-1, -3)$ gap as a function of the magnetic field. For the $1.06^\circ$ sample, the critical magnetic field for this gap is determined by the crossing point of the $(-K,\downarrow)$ and $(-K,\uparrow)$ Hofstadter spectra, occurring at approximately $9.5$~T (see Fig.~\ref{fig3} (b)). In contrast, there is no such Zeeman induced crossing for the $(-1, -3)$ state in the $1.03^\circ$ sample. Its gap decreases rapidly at the rational flux $\Phi/\Phi_0 = 1/3$ ($8.5$~T), while a very small gap persists and remains observable up to about $B \approx 7$~T (see Fig.~\ref{fig3} (c)). This behavior is in good agreement with the experimental results.
Note that the energy resolution in the calculation is set to the experimental value of approximately $0.02$~meV.  Interestingly, it implies that when the Fermi energy lies within the (-1, -3) gap, the two samples are in distinct flavor-polarized states. 

Our model can, in fact, effectively interpret a series of other experiments, and the relevant calculations are provided in the Supplementary Material. The reason this model can so effectively explain numerous experiments lies in the fact that the Wannier diagram essentially describes the topological nature of the Hofstadter spectrum, which depends only on the ordering of the Hofstadter states and not on the specific sizes of the energy gaps. The Zeeman splitting terms capture the physical essence, namely the flavor-polarized Hofstadter spectrum that varies with the magnetic field, thereby correctly describing the ordering of Hofstadter states in MATBG.
Despite this success, our model cannot yet fully quantitatively predict the size of each energy gap. It may originate from factors such as filling-dependent effective $g$-factors and the intricate uncertainties of experimental measurements.

\emph{Further predictions}---Finally, we predict that applying an in-plane magnetic field \( B_{\parallel} \) can modulate the Zeeman splitting, while keeping the valley splitting unchanged, thereby representing an effective knob to control the crossings between flavor-polarized Hofstadter spectra and significantly altering the CCI states in Wannier diagram. In Fig.~\ref{fig3} (d), we take the 1.06\(^\circ\) sample as an example to demonstrate the  effects of \( B_{\parallel} \) on the \( (-1,-3) \) gap in Hofstadter spectrum. In Fig.~\ref{fig3} (e), we further plot the \( (-1,-3) \) gap as a function of magnetic filed under various \( B_{\parallel} \). It reveals that \( B_{\parallel} \) can substantially reduce the critical magnetic field at which the \( (-1,-3) \) gap emerges. These predictions can serve as a further verification of our model.

\emph{Summary}.---To summarize, we distill the core conclusions from our Hofstadter spectrum model of MATBG into the following points:

\begin{enumerate}[label=(\arabic*), leftmargin=*, nosep, before=\vspace{0.5\baselineskip}, after=\vspace{0.5\baselineskip}]
    \item Our model can essentially be regarded as a mean-field description of the CCI states in MATBG, with the key insight that the primary effect of correlations is to enhance the spin and valley Zeeman effects via correlation-enhanced $g$-factors~\cite{lu2019superconductors,jiang2019valley,vallejo2021detection,PhysRevB.102.121406,du2009fractional,song2010high,jeong2024interplay,PhysRevLett.96.136806,PhysRevLett.108.106804,PhysRevLett.124.097601,PhysRevLett.99.106802,PhysRevB.100.085437,ren2020spectroscopic}.
    
    \item In MATBG, the extremely narrow bandwidth of the moir\'e flat bands makes the Zeeman effect indispensable to describe the Hofstadter spectrum. The striking consistency between our model and experiments confirms it captures the core physics: a Zeeman-induced flavor-polarized Hofstadter butterfly, shaped by Zeeman-driven intertwining.
    
    
    \item Our model fails for very small magnetic fields, where the Zeeman splittings are small and other correlation effects dominate. And it also cannot account for the interacting Hofstadter states with fractional $t$ or $s$ values, which may require more advanced techniques.
    
\end{enumerate}
At last, we mention a key difference between the Hofstadter spectrum of the MATBG and the TMD moiré structures. In TMD systems, strong spin–orbit coupling enforces spin–valley locking, so that its Hofstadter spectrum always has a two-flavor-polarized spectrum~\cite{zhao2025hofstadter,foutty2025magnetic,wang2024phase}. In contrast,  the spin and valley are decoupled in MATBG, thus exhibiting four flavor-polarized spectra.    
\begin{acknowledgments}
    This work was supported by the National Key Research and Development Program of China (Grants No.~2022YFA1403501), the National Natural Science Foundation of China (Grants No.12474169, No.12574043, and No.12174130).
\end{acknowledgments}

\bibliography{references}

\clearpage
\onecolumngrid

\newcommand{\bk}{\bm{k}}
\newcommand{\bq}{\bm{q}}
\newcommand{\btk}{\widetilde{\bm{k}}}
\newcommand{\btq}{\widetilde{\bm{q}}}
\newcommand{\br}{\bm{r}}
\newcommand{\cop}{\hat{c}}
\newcommand{\dop}{\hat{d}}
\newcommand{\xmark}{\ding{55}}
\def\Red#1{\textcolor{red}{#1}}
\def\Blue#1{\textcolor{blue}{#1}}
\setcounter{section}{0}
\renewcommand{\thesection}{S\arabic{section}}
\setcounter{secnumdepth}{1}

\begin{center}
\textbf{\large Supplemental Information for “Theory of Correlated Hofstadter Spectrum in Magic-Angle Graphene"}

\vspace{0.4cm}

{\normalsize
Chen Zhao$^{1}$,
Zhaowen Miao$^{1}$,
Zhen Ma$^{2}$,
Ying-Hai Wu$^{1,*}$,
Ming Lu$^{3,\dagger}$,
Jin-Hua Gao$^{1,\ddagger}$,
and X. C. Xie$^{4,5,6}$
}

\vspace{0.25cm}

{\small
\textit{$^{1}$School of Physics and Wuhan National High Magnetic Field Center,\\ Huazhong University of Science and Technology, Wuhan, Hubei 430074, China\\
$^{2}$School of Electronics and Information, Zhengzhou University of Light Industry, Zhengzhou, 450002 China\\
$^{3}$Beijing Academy of Quantum Information Sciences, Beijing 100193, China\\
$^{4}$International Center for Quantum Materials, School of Physics, Peking University, Beijing 100871, China\\
$^{5}$Institute for Nanoelectronic Devices and Quantum Computing, Fudan University, Shanghai 200433, China\\
$^{6}$Hefei National Laboratory, Hefei 230088, China\\[0.2cm]
}}

\vspace{0.6cm}
\end{center}

\setcounter{equation}{0}
\setcounter{figure}{0}
\setcounter{table}{0}
\setcounter{page}{1}
\makeatletter
\renewcommand{\theequation}{S\arabic{equation}}
\renewcommand{\thefigure}{S\arabic{figure}}
\renewcommand{\bibnumfmt}[1]{[S#1]}

\section{Band Structure of MATBG with Substrate-Induced Gaps}

\begin{figure*}[h]
\centering
\includegraphics[width=18cm]{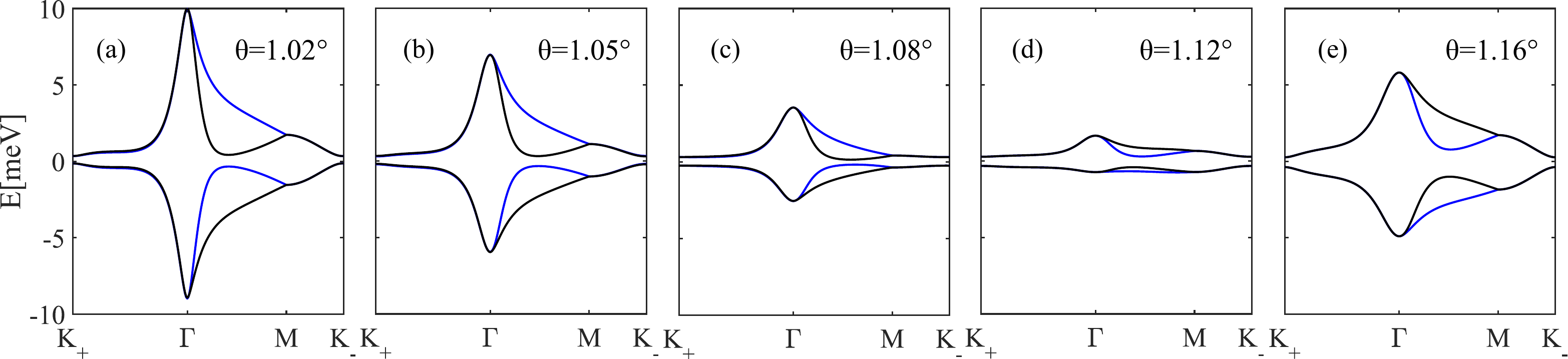}
\caption{Single-particle band structures of MATBG at twist angles (a) $\theta = 1.02^\circ$, (b) $\theta = 1.05^\circ$, (c) $\theta = 1.08^\circ$, (d) $\theta = 1.12^\circ$, and (e) $\theta = 1.16^\circ$, calculated within the BM model including a small substrate-induced mass term $\Delta = 1~\text{meV}$.}
\label{figs1}
\end{figure*}

In this section, we summarize the single-particle band structure of MATBG used in our calculations and highlight two independent aspects that are key to determining the Hofstadter spectrum: a small substrate-induced mass term and the twist-angle dependence of the flat-band bandwidth.

We include a mass term $\Delta\sigma_z$ in the BM model to describe the effect of hBN substrates. We set $\Delta = 1$~meV, inducing a gap of about $0.5$~meV at charge neutrality in each valley. Although this gap is too small to create a large transport gap, it is comparable to the bandwidth of the lowest moiré bands and endowing the two valleys with distinct orbital magnetic moments\cite{PhysRevB.81.195431}. Consequently, it lifts the valley degeneracy under magnetic field that is essential for obtaining the correct energy ordering of the flavor-polarized Hofstadter subbands. Without this gap, the two valleys would be degenerate under magnetic field.

The twist angle has a significant impact on the bandwidth of the low-energy moiré flat bands. As shown in Fig.~\ref{figs1}(a)–(e), the bands near charge neutrality are narrowest at $\theta = 1.12^\circ$, and become broader at other twist angles. This bandwidth variation significantly impacts the Hofstadter spectrum by changing the intersection between flavor-polarized Hofstadter spectrum, reshaping the hierarchy of Hofstadter gaps, and modifying the appearance and critical fields of the correlated Chern insulator states discussed in the main text.

\section{The Calculation of the Hofstadter Spectrum}

We calculate the Hofstadter spectrum of the MATBG using the method described in Ref.~\cite{bistritzer2011moire,hejaz2019landau}. We work with the BM model described in Eq.~2 of the main text. It should be noted that here we first ignore the valley and spin Zeeman terms and add them back later. We use Landau gauge $\mathbf{A}=(-yB_z,0)$, and diagonalize the Hamiltonian in the Landau basis. The raising and lowering operators of the Landau levels index are defined as 
\begin{equation}
\begin{aligned}
    {\frac{\sqrt{2}}{\ell_B}}a&=-\partial_y+k_x+\frac{\sqrt{3}}{2}k_\theta-\frac{eB_zy}{\hbar} \\
    \frac{\sqrt{2}}{\ell_B}a^\dagger&=\partial_y+k_x+\frac{\sqrt{3}}{2}k_\theta-\frac{eB_zy}{\hbar}
\end{aligned}
\end{equation}

The intralayer part of $H_0$ is given by
\begin{equation}
    H_\mathrm{intra}=\hbar v_F\tau_z\frac{(\eta_x+i\tau_z\eta_y)}{2}(\frac{\sqrt{2}}{\ell_B}a+i\eta_z\frac{k_\theta}{2})+\mathrm{H.c.}+\Delta\sigma_z
\end{equation}
According to Eq.~9 of the main text, the interlayer terms couple the Landau basis with different values of $k_x\pm \lambda \frac{\sqrt{3}}{2}k_\theta$. When the phase change induced by the guiding center change satisfies $\frac{3\sqrt{3}}4k_\theta^2\ell_B^2=2\pi\frac{p}{q}$, the guiding center $y_c=k_x\ell_B^2$ becomes periodic. In this case, the magnetic flux per moir\'e unit cell satisfies $\Phi/\Phi_0=q/2p$.

Then we can write the guiding center as $y_c=y_1(mq+j)\frac{\sqrt{3}}{2}k_\theta \ell_B^2$, where $m$ is a integer and $j=0,1,...,q-1$. We then conduct a Fourier transform with respect to $m$. The resulting magnetic Brillouin zone is 
\begin{equation}
    \{(k_1,k_2)|0<k_1<\frac{\sqrt{3}}{2}k_\theta,0<k_2<\frac{3k_\theta}{2p} \}
\end{equation}

We then construct the Hamiltonian in the new basis $\ket{l,n,\sigma,k_1,k_2,j}$. To avoid cutoff effects, we include $n_\text{max}=[20(\text{max}(\hbar v_Fk_\theta,w)/\omega_c)^2]$ Landau levels, where $w=110\text{meV}$ and $\omega_c=\sqrt{2}\hbar v_F/\ell_B$. It is important to note that for the valley \(K\), the \(n\)th Landau level on sublattice A couples to the \((n+1)\)th Landau level on sublattice B, while the opposite holds for the valley \(-K\). Therefore, we set the cutoff indices of the two sublattices differ by one to avoid an artificial zero-energy state. The intralayer terms can be written as 
\begin{equation}
\begin{aligned}
    H_\mathrm{intra}&=\hbar v_F\tau_z\frac{(\eta_x+i\tau_z\eta_y)}{2}(\sqrt{\frac{3\sqrt{3}}{4\pi}\frac{q}{p}}a+\frac{i\eta_z}{2})+\mathrm{H.c.}+\Delta\sigma_z \\
\end{aligned}
\end{equation}

The interlayer coupling is
\begin{equation}
    \begin{aligned}
        \bra{1,n^\prime,\sigma^\prime,k_1^\prime,k_2^\prime,j^\prime}T_0e^{-i\tau_z\mathbf{q}_0\cdot\mathbf{r}}\ket{2,n,\sigma,k_1,k_2,j}&=\delta_{n,n^\prime}\delta_{k_1,k_1^\prime}\delta_{k_2,k_2^\prime}\delta_{j,j^\prime}\bra{1,\sigma^\prime}T_0\ket{2,\sigma}, \\
        \bra{1,n^\prime,\sigma^\prime,k_1^\prime,k_2^\prime,j^\prime}T_1e^{-i\tau_z\mathbf{q}_1\cdot\mathbf{r}}\ket{2,n,\sigma,k_1,k_2,j}&=\delta_{n,n^\prime}\delta_{k_1,k_1^\prime}\delta_{k_2,k_2^\prime}\delta_{(j+\xi),j^\prime}F_{n,n^\prime}(\tau_z z_1)\\
        &e^{-i\tau_z\frac{3}{2}k_\theta k_1\ell_B^2}e^{-i\tau_z\frac{\sqrt{3}}{2}k_\theta k_2\ell_B^2}e^{-i\tau_z\frac{2\pi p}{q}(j+\frac{1}{2})}\bra{1,\sigma^\prime}T_1\ket{2,\sigma}, \\
        \bra{1,n^\prime,\sigma^\prime,k_1^\prime,k_2^\prime,j^\prime}T_2e^{-i\tau_z\mathbf{q}_2\cdot\mathbf{r}}\ket{2,n,\sigma,k_1,k_2,j}&=\delta_{n,n^\prime}\delta_{k_1,k_1^\prime}\delta_{k_2,k_2^\prime}\delta_{(j-\xi),j^\prime}F_{n,n^\prime}(\tau_z z_2)\\
        &e^{-i\tau_z\frac{3}{2}k_\theta k_1\ell_B^2}e^{i\tau_z\frac{\sqrt{3}}{2}k_\theta k_2\ell_B^2}e^{-i\tau_z\frac{2\pi p}{q}(j-\frac{1}{2})}\bra{1,\sigma^\prime}T_2\ket{2,\sigma}
    \end{aligned}
\end{equation}

\begin{figure*}[h]
\centering
\includegraphics[width=10cm]{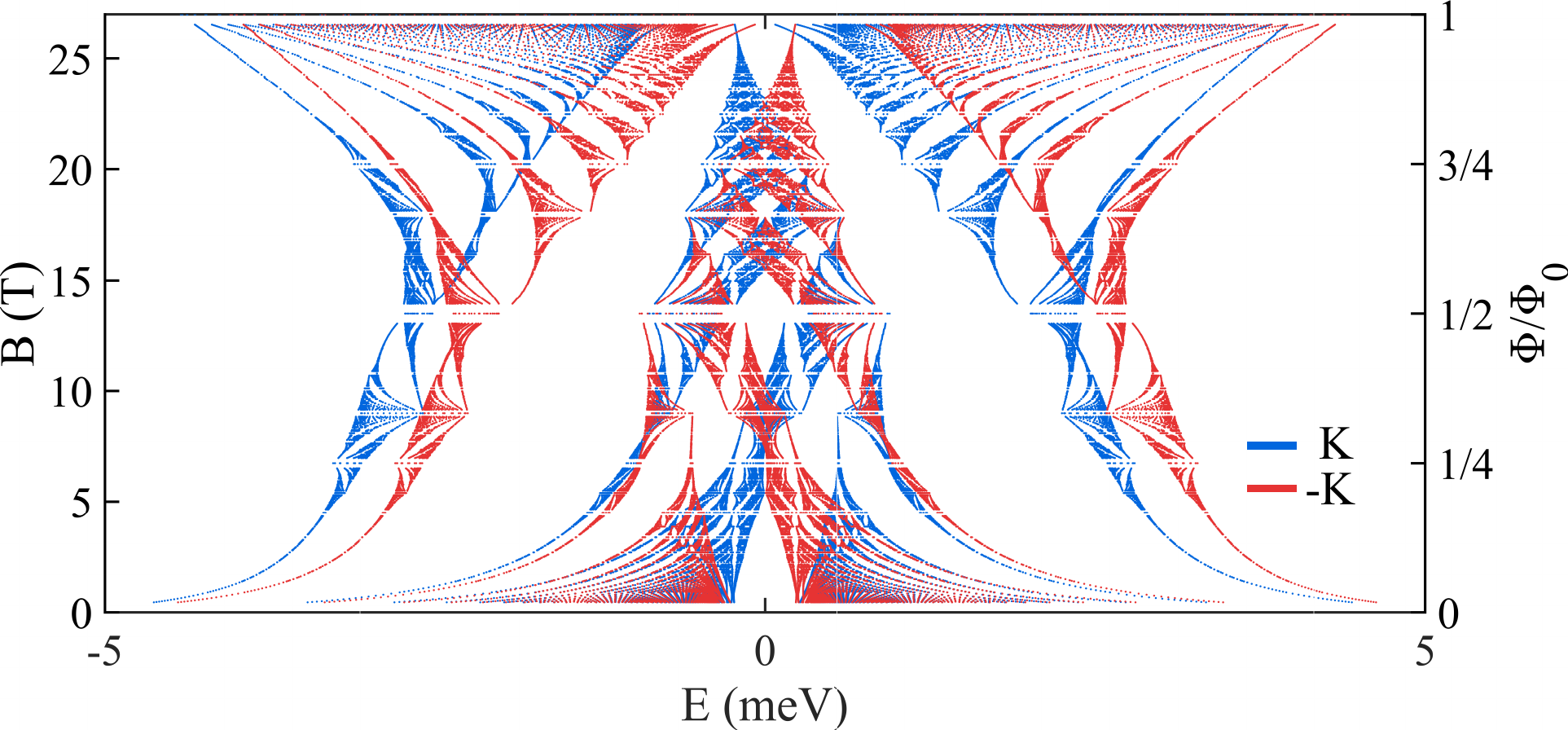}
\caption{Hofstadter spectra of MATBG at $\theta = 1.06^\circ$ with a substrate-induced mass term $\Delta = 1~\text{meV}$ while neglecting the valley and spin Zeeman terms. Without this mass term, the spectra of the two valleys would be degenerate under the magnetic field.}
\label{figs2}
\end{figure*}

\section{The Influence of the Mass Term}

In graphene-based Dirac systems, a sublattice mass term opens a gap at zero field and, in a perpendicular magnetic field, causes the zeroth Landau levels in the two valleys to shift in opposite directions, thereby lifting the valley degeneracy~\cite{PhysRevB.81.195431}. In our calculations, this shift significantly reconstructs the Hofstadter spectrum, as illustrated in Fig.~\ref{figs2}. This effect is essential for reproducing the experimentally observed Wannier diagrams.

\section{Additional examples for the two gap-closing mechanisms}

\begin{figure*}[h]
\centering
\includegraphics[width=18cm]{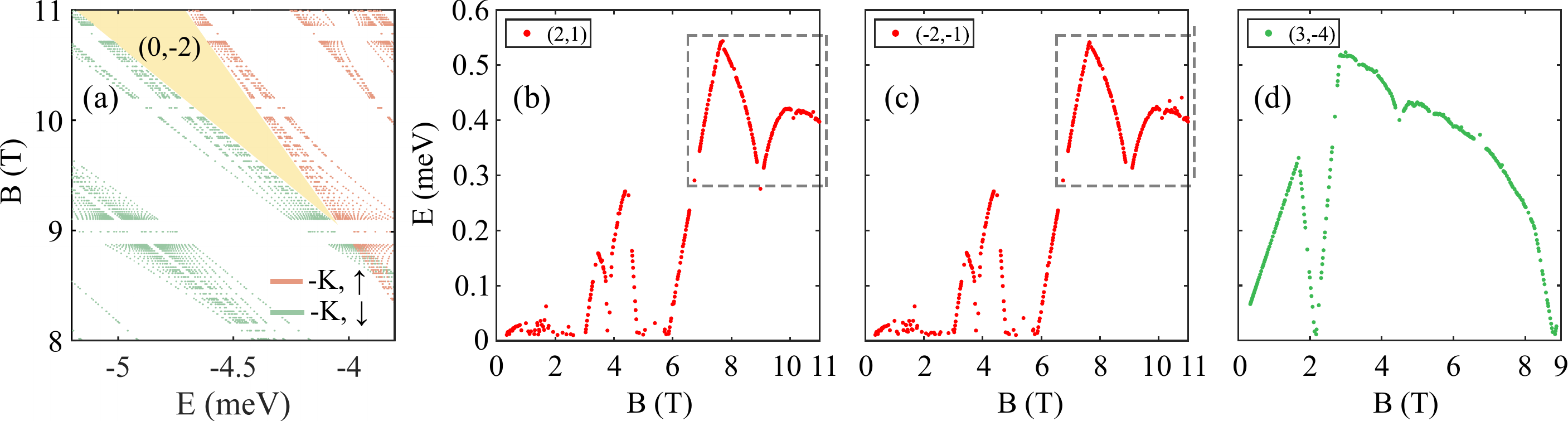}
\caption{(a) is the magnified version of the Hofstadter spectrum with $\theta=1.06^\circ$. The yellow shaded region marks the gap of the $(0,-2)$ state. (b) and (c) are the evolution of gaps of the $(2,1)$ and $(-2,-1)$ states with $\theta=1.06^\circ$. The dashed box denotes the experimentally observed region. (d) is the evolution of gaps of the $(3,-4)$ state with $\theta=1.06^\circ$, indicating that this gap is interrupted at $\Phi/\Phi_0=1/3~(9~\mathrm{T})$.}
\label{figs3}
\end{figure*}

Fig.~\ref{figs3} shows some examples of the two gap-closing mechanisms introduced in the main text. In panel~(a), the $(0,-2)$ gap corresponds to the large gap between the
 flavor-polarized Hofstadter spectra of $\ket{-K,\downarrow}$ and $\ket{-K,\uparrow}$. Due to their intersection, this gap is interrupted around $B \approx 9\,\mathrm{T}$. In panels~(b) and (c), the $(2,1)$ and $(-2,-1)$ gaps both reach their maximal values in the field range $B \approx 6.7$--$11\,\mathrm{T}$, which coincides with the region observed in experiment. At $B = 6.7\,\mathrm{T}$ ($\Phi/\Phi_0 = 1/4$), these gaps are interrupted and rapidly shrink as the magnetic field decreases. The $(3,-4)$ gap in panel~(d) provides another clear example of the rational-flux mechanism: consistent with the experimental observations~\cite{yu2022correlated}, this gap is interrupted at $\Phi/\Phi_0 = 1/3$ ($9\,\mathrm{T}$).

\section{Comparison with the Experimental Results}

In this section, we compare representative experimental observations with our calculations across several twist angles and field ranges. All simulations use exactly the same model and parameters as in the main text.

\subsection{$\theta=1.08^\circ$ -- Comparison with experiment in Ref.~\cite{PhysRevLett.127.197701}}

\begin{figure*}[h]
\centering
\includegraphics[width=18cm]{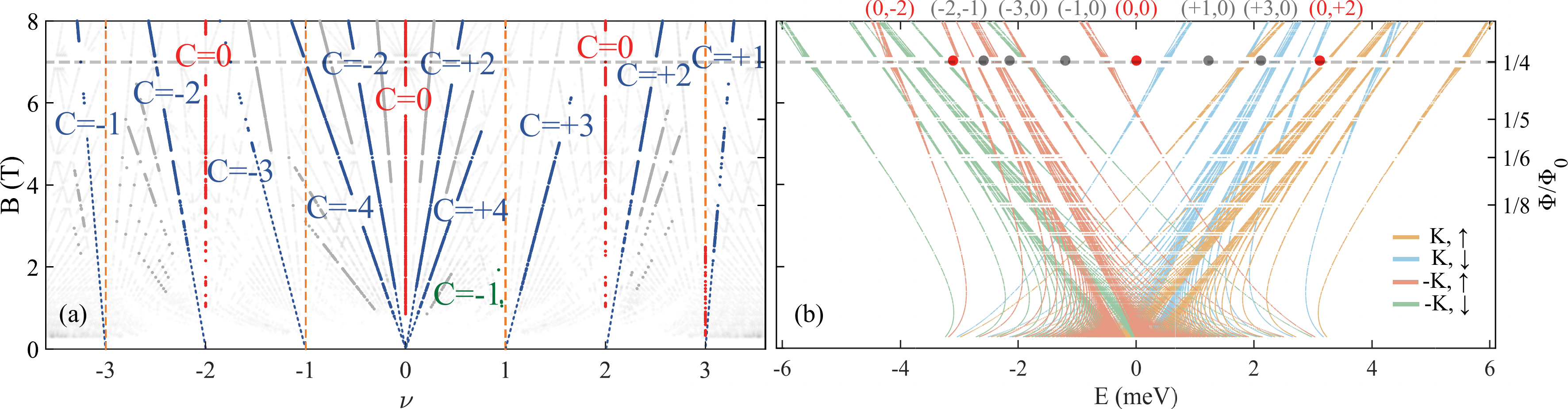}
\caption{Hofstadter spectrum of MATBG with twist angle $\theta=1.08^\circ$. (a) Wannier diagram. The observed CCI states in experiment are highlighted with the same color scheme as that in Ref.~\cite{PhysRevLett.127.197701}. (b) Flavor-polarized Hofstadter spectra. The minimum energy resolution for gaps is $0.01$ meV. All the other parameters are the same as Fig.~1 of the main text.}
\label{figs4}
\end{figure*}

We compare our results with the transport experiment of Stepanov et al. (Ref.~\cite{PhysRevLett.127.197701}), which investigated a MATBG device with twist angle $\theta \approx 1.08^\circ$ at a base temperature $T \approx 50~\mathrm{mK}$. As shown in Fig.~\ref{figs4}, (a) presents the Wannier diagram with color coding following Ref.~\cite{PhysRevLett.127.197701}, and (b) shows the corresponding Hofstadter spectrum with several major gaps denoted by their $(t,s)$ indices. The experiment combined zero-field and finite-field transport and Hall measurements. For our modeling at the same twist angle $\theta = 1.08^\circ$, we identify Hofstadter gaps in the Wannier diagram using a minimum energy resolution of $0.01~\mathrm{meV}$. At this angle, the flat-band bandwidth is slightly smaller than that at $\theta = 1.06^\circ$, as shown in Fig.~\ref{figs1}(c). This leads to modest shifts in flavor-resolved crossings and the visibility hierarchy of CCI gaps. Unlike the $\theta = 1.06^\circ$ case in the main text, Stepanov et al.\ clearly resolve vertical $C = 0$ lines at $\nu \approx \pm 2$, which is consistent with our calculations. This enhanced visibility may be facilitated by the lower experimental temperature. These vertical $C = 0$ lines correspond to gaps between different flavor-resolved Hofstadter subbands.

\subsection{$\theta=1.12^\circ$ -- Comparison with experiment in Ref.~\cite{saito2021hofstadter}}

\begin{figure*}[h]
\centering
\includegraphics[width=18cm]{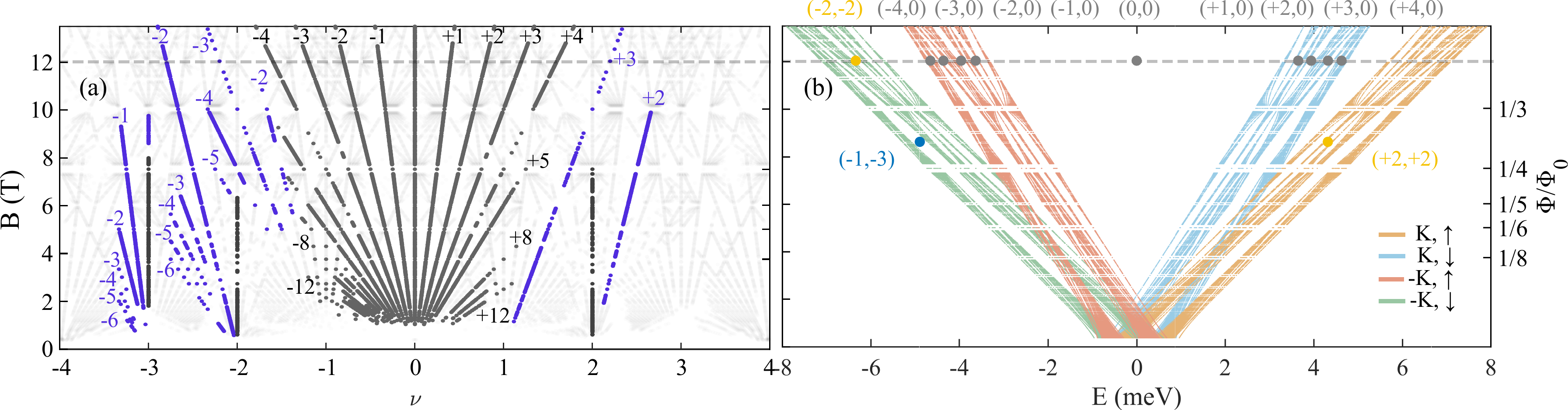}
\caption{Hofstadter spectrum of  MATBG with twist angle $\theta=1.12^\circ$. (a) Wannier diagram. The observed CCI states in experiment are highlighted with the same color scheme as that in Ref.~\cite{saito2021hofstadter}. (b) Flavor-polarized Hofstadter spectra. The minimum energy resolution for gaps is $0.01$ meV. All the other parameters are the same as Fig.~1 of the main text.}
\label{figs5}
\end{figure*}

We compare our results with the transport experiment of Saito et al. (Ref.~\cite{saito2021hofstadter}), which investigated a MATBG device with twist angle $\theta \approx 1.12^\circ$ at a base temperature $T \approx 10~\mathrm{mK}$. As shown in Fig.~\ref{figs5}, (a) presents the Wannier diagram with color coding following Ref.~\cite{saito2021hofstadter}, and (b) shows the corresponding Hofstadter spectrum with several major gaps denoted by their $(t,s)$ indices. In comparison with the $\theta = 1.06^\circ$ case, $\theta = 1.12^\circ$ is the magic angle under our parameter set, yielding an exceptionally narrow Hofstadter bandwidth, as shown in Fig.~\ref{figs1}(d). Consequently, by $B \approx 7.5~\mathrm{T}$ ($\Phi/\Phi_0 = 1/4$), the four flavor-resolved Hofstadter spectra are fully separated, driving a pronounced reorganization of gaps. A key difference is that the $(t,s) = (-1,-3)$ gap becomes an inter-flavor magnetic subband gap. It is larger and more easily resolved at low fields, but weakens and eventually closes at higher fields, which is in agreement with the experimental observations.

\subsection{$\theta=1.17^\circ$ -- Comparison with experiment in Ref.~\cite{wu2021chern}}

\begin{figure*}[h]
\centering
\includegraphics[width=18cm]{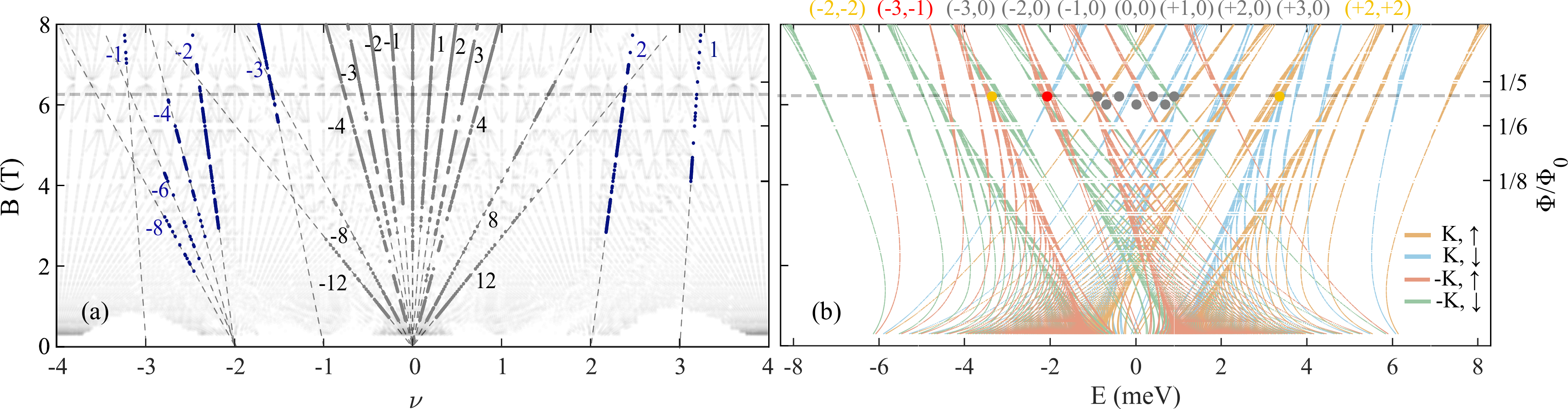}
\caption{Hofstadter spectrum of  MATBG with twist angle $\theta=1.17^\circ$. (a) Wannier diagram. The observed CCI states in experiment are highlighted with the same color scheme as that in Ref.~\cite{wu2021chern}. (b) Flavor-polarized Hofstadter spectra. The minimum energy resolution for gaps is $0.01$ meV. All the other parameters are the same as Fig.~1 of the main text.}
\label{figs6}
\end{figure*}

We compare our results with the transport experiment of Wu et al. (Ref.~\cite{wu2021chern}), which investigated a MATBG device with twist angle $\theta \approx 1.17^\circ$ at a base temperature $T \approx 300~\mathrm{mK}$. As shown in Fig.~\ref{figs6}, (a) presents the Wannier diagram with color coding following Ref.~\cite{wu2021chern}, and (b) shows the corresponding Hofstadter spectrum with several major gaps denoted by their $(t,s)$ indices. Compared with the $\theta = 1.06^\circ$ case, the $\theta = 1.17^\circ$ device exhibits a slightly larger bandwidth, as shown in Fig.~\ref{figs1}(e). This leads to more intricate overlap among flavor-resolved Hofstadter spectra. Together with plausible influences from temperature and measurement conditions, fewer Hofstadter states are resolved.

\subsection{$\theta=1.06^\circ$ -- Comparison with experiment in Ref.~\cite{das2021symmetry}}

\begin{figure*}[h]
\centering
\includegraphics[width=18cm]{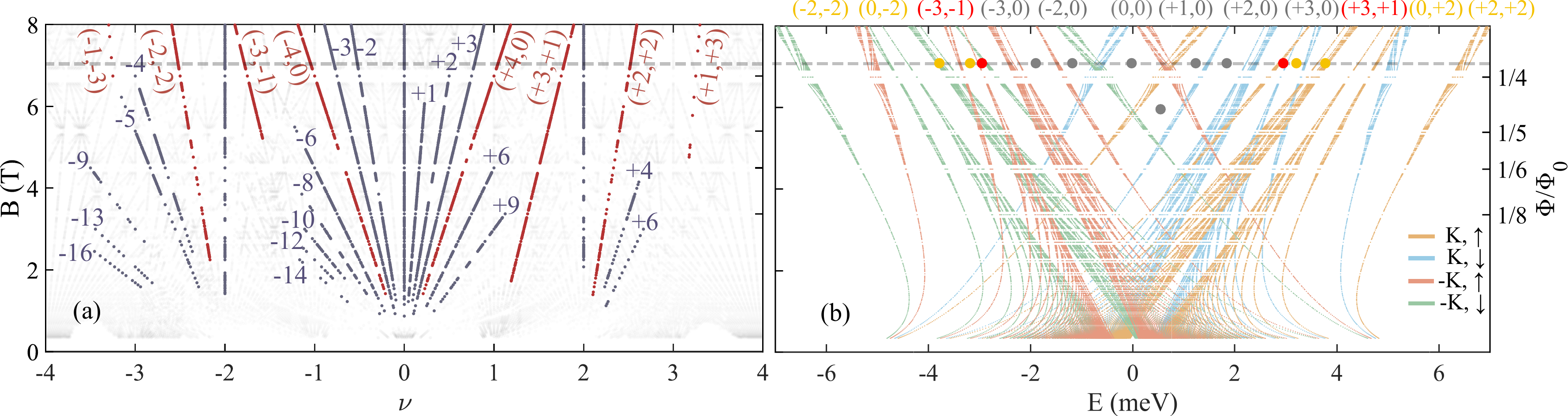}
\caption{Hofstadter spectrum of  MATBG with twist angle $\theta=1.06^\circ$. (a) Wannier diagram. The observed CCI states in experiment are highlighted with the same color scheme as that in Ref.~\cite{das2021symmetry}. (b) Flavor-polarized Hofstadter spectra. The minimum energy resolution for gaps is $0.01$ meV. All the other parameters are the same as Fig.~1 of the main text.}
\label{figs7}
\end{figure*}

We compare our results with the transport experiment of Das et al. (Ref.~\cite{das2021symmetry}), which investigated a MATWBG device with twist angle $\theta \approx 1.06^\circ$ at a base temperature $T \approx 30~\mathrm{mK}$. As shown in Fig.~\ref{figs7}, (a) presents the Wannier diagram with color coding following Ref.~\cite{das2021symmetry}, and (b) shows the corresponding Hofstadter spectrum with several major gaps denoted by their $(t,s)$ indices. Notably, the observed $(-1, -3)$ state differs from Fig.~1 of the main text. And additional high |t| trajectories are also resolved in the experiment. We attribute the differences to variations in measurement temperature and sample quality. The experiment in Fig.~\ref{figs6} was performed at $T\approx30$ mK~\cite{das2021symmetry}, with an energy resolution roughly an order of magnitude lower than that of Fig.~1 of the main text. It thus can resolve some states that are not visible in Fig.~1. Overall, these differences are consistent with our interpretation, and flavor-resolved crossings are sensitive to temperature and disorder.

\end{document}